\documentclass[12pt]{iopart}

\usepackage{amsfonts}
\usepackage{graphicx}
\newcommand{\be}{\begin{equation}}
\newcommand{\ee}{\end{equation}}

\newcommand{\bea}{\begin{eqnarray}}
\newcommand{\eea}{\end{eqnarray}}

\begin{document}

\title[Poisson algebra of quasilocal angular momentum and its asymptotic limit]{Poisson algebra of quasilocal angular momentum and its asymptotic limit}
\author{Jong Hyuk Yoon$^1$ and Seung Hun Oh$^2$}
\address{School of Physics, Konkuk University, 120 Neungdong-ro, 
Gwangjin-gu, Seoul 05029, Korea} 
\vspace{-6pt}
\ead{$^1$yoonjh3404@gmail.com, $^2$shoh.physics@gmail.com}

\begin{indented}
\item[]December 2016
\end{indented}

\begin{abstract} 
We study the previously proposed quasilocal angular momentum of gravitational fields in the absence of isometries. The quasilocal angular momentum $L(\xi)$ has the following attractive properties; ({\it i}) it follows from the Einstein's constraint equations,
({\it ii}) it satisfies the Poisson algebra
$\{L(\xi), L(\eta) \}_{\rm P.B.} = ({1/16\pi)}\, L( [\xi, \eta]_{\rm L} )$, ({\it iii}) its Poisson algebra reduces to the standard $SO(3)$ algebra of angular momentum at null infinity, and ({\it iv}) it reproduces the standard value for the Kerr spacetime at null infinity.
It will be argued that our definition is a quasilocal and canonical generalization of A. Rizzi's geometric definition at null infinity.
We also propose a new definition of an {\it invariant} quasilocal angular momentum $L^{2}$ 
such that
\begin{math}
\{ L^{2}, L(\xi)  \}_{\rm P.B.}  = 0,
\end{math}  
which becomes $(ma)^{2}$ at the null infinity of the Kerr spacetime.
Therefore, it may be regarded 
as a quasilocal generalization of the Casimir invariant of ordinary angular momentum in the flat spacetime.

\end{abstract}

%%%%%%%%%%%%%%%%%%%%%%%%%%%%%%%%%%%%%%%%%%%%%%%%%%%%%%%%%%%%%%%%%%%%%%%%%%%%%%%%%%%%%%%

\section{Introduction}  \label{intro}
Einstein's theory of general relativity is based on the general covariance and the equivalence principle. This implies that local definitions of non-zero gravitational energy, linear and angular momentum are incompatible with the general covariance. However, as is well-known, there have been several attempts 
to define local gravitational energy and momentum. Einstein himself introduced energy-momentum pseudotensor of gravitational fields, which was further developed by L. Landau and E. Lifshitz \cite{Landau}. In these approaches, local definitions of gravitational energy-momentum and the corresponding conservation laws were found, but only at the cost of the general covariance. 

It is well-known that the Einstein's theory is a highly constrained system with four non-linear constraints enforcing the general covariance \cite{teitelboim74}. By analyzing these constraint equations, R. Arnowitt, S. Deser, and C.W. Misner (ADM) defined canonical
gravitational energy-momentum as symmetry generators of the global Poincar{\'e} group at spatial infinity \cite{misner62}. The quasilocal generalizations of gravitational energy-momentum on a two-sphere at a finite distance on spacelike hypersurfaces were proposed by several authors \cite{penrose82,brown_york92,brown_york93,brown_york99,nester99,nester11,szabados09}, but it is highly non-trivial to extract dynamical contents such as the in- and out-going fluxes of gravitational energy-momentum from these expressions \cite{dray85,nester_flux}.

On the other hand, H. Bondi, A.W.K. Metzner, and R.K. Sachs (BMS) defined gravitational energy-momentum at null infinity, whose difference at two different retarded times naturally defines the out-going gravitational flux \cite{BVM,sachs,ludvigsen83}. The definition of gravitational angular momentum remains tricky, however, as there are no natural notions of the center of rotation as seen by an observer at null infinity \cite{moreschi86,szabados_angular99,epp2000}.

To our knowledge, the most successful definition of gravitational angular momentum at null infinity is given by A. Rizzi \cite{rizzi98}.  He defined the angular momentum of gravitational fields as a two-surface integral of the inner product of the $SO(3)$ generators 
and the commutator of the in- and out-going null vector fields at null infinity. This definition is geometrical and reproduces the standard value for the Kerr spacetime. Despite its successful features, Rizzi's definition has several
drawbacks; it is gauge-dependent in that the null vectors that play crucial roles in his definition must be affinely parametrized \cite{christodoulou}, and it is not expressed in canonical variables, and moreover, it is not at all clear whether it is compatible with the Einstein's constraints. 

A list of guidelines of a satisfactory quasilocal generalization of gravitational angular momentum would be the followings \cite{penrose82,szabados09,rizzi98};
(\textit{i}) it must be a two-surface integral of some scalar density function on the gravitational phase space,
(\textit{ii}) it must be independent of how the two-surface in a given spacetime is labeled,
(\textit{iii}) it must incorporate the anti-symmetry intrinsic to the angular momentum, 
(\textit{iv}) it must be interpretable geometrically, 
(\textit{v}) the in- and/or out-going quasilocal fluxes of angular momentum must be given by the time-derivative of the quasilocal angular momentum, 
(\textit{vi}) it must reproduce the standard value of angular momentum for the Kerr spacetime in the asymptotic limit,
(\textit{vii}) its algebra must reduce to the angular momentum algebra 
$[L_i , L_j]= -{\epsilon_{ij}}^k L_k$ in the asymptotic limit, 
(\textit{viii}) finally, though not necessary, it is certainly desirable that the quasilocal gravitational angular momentum be deducible from the Einstein's constraints.

In this paper, we study in detail the quasilocal gravitational angular momentum that meets the above criteria, which was proposed years ago \cite{yoon04}. However, such quasilocal angular momentum depends on a divergence-free tangent vector field on a compact two-surface.
It is well-known that, by the Poincar{\'{e}}-Hopf theorem, any continuous tangent vector field has at least one zero for any compact regular even dimensional manifold with non-zero Euler characteristic.
Therefore, our quasilocal angular momentum fails to ``measure'' the rotation of gravitational fields at these singular points of the divergence-free vector field.
Thus, it is desirable to have an invariant notion of 
quasilocal gravitational angular momentum, as a gravitational analog of {\it Casimir invariant} of ordinary angular momentum, which has the following properties;
(\textit{i}) it must commute with quasilocal angular momentum under the Poisson bracket, 
(\textit{ii}) it must be independent of the divergence-free vector field,
(\textit{iii}) it must reproduce $(ma)^{2}$ at null infinity of the Kerr spacetime.
By the property (\textit{ii}), the invariant quasilocal angular momentum will be insensitive to the singular nature of the vector field. We will propose a two-surface integral that defines an {\it invariant} quasilocal angular momentum with these properties.

In Section \ref{background}, we summarize (2+2) canonical formalism of Einstein's theory and present the Einstein's equations. Especially, it will be seen that the integral forms of the constraint equations are precisely the quasilocal balance equations of energy, linear momentum, and angular momentum of gravitational 
fields \cite{ashtekar02}. 
In Section \ref{gauge}, we show that the sufficient and necessary condition that
quasilocal angular momentum $L(\xi)$ of a two-surface along a given 
vector field $\xi^a$ is independent of the ways of labeling the surface is
\bea
\tilde{\nabla}_a \xi^a = 0,             \label{taste}
\eea
where $\tilde{\nabla}_a  $ is the covariant derivative on two-surface.
In Section \ref{geometric}, geometric interpretations of the quasilocal energy and linear momentum will be presented and the reference terms of these quasilocal quantities are discussed. 
In Section \ref{poisson}, we will focus on the flux aspects of quasilocal angular momentum and prove a theorem that the 3-dimensional flux integral representations of quasilocal angular momentum associated with the vector fields $\xi$ and $\eta$ are closed under the Poisson bracket \cite{teitelboim_book,yoon14}:
\bea
\{ L(\xi), L(\eta) \}_{\rm P.B.} ={1\over 16\pi} L([\xi, \eta]).   \label{juice}
\eea
In Section \ref{invariant}, we introduce an {\it invariant} quasilocal angular momentum $L^{2}$ that  commutes with  $L(\xi)$ for any divergence-free vector field $\xi^{a}$,  
\be
 \{ L^{2}, L(\xi)  \}_{\rm P.B.}  = 0.  \label{monitor}
\ee 
In Section \ref{asymptotic}, we examine properties of our quasilocal angular momentum 
$L(\xi)$ and invariant quasilocal angular momentum $L^{2}$ at null infinity. 
Moreover, it is shown that the Poisson algebra of $L(\xi)$ reduces to the $SO(3)$ Poisson algebra of ordinary angular momentum at null infinity.
In Section \ref{rizzi}, it is shown that Rizzi's geometric angular momentum at null infinity is the limiting form of our quasilocal angular momentum. Thus, our angular momentum may be regarded 
as a quasilocal generalization of Rizzi's definition at null infinity \cite{rizzi98}. 
In Section \ref{discussion}, we summarize and discuss other notable aspects of our quasilocal angular momentum.

%%%%%%%%%%%%%%%%%%%%%%%%%%%%%%%%%%%%%%%%%%%%%%%%%%%%%%%%%%%%%%%%%%%%%%%%%%%%%%%%%%%%%%%%%%%%

\section{Background}  \label{background}
In this section we will review the (2+2) formalism of Einstein's theory \cite{yoon04}. In this formalism, the 4-dimensional spacetime $E_4$ is regarded as a fibre bundle that consists of a 2-dimensional base space $M_{1+1}$ of the Lorentzian signature and a 2-dimensional spacelike fibre $N_2$ at each point on $M_{1+1}$. Then, the Einstein's theory can be interpreted as a gauge theory
defined on a (1+1)-dimensional base manifold with {\it diff}$(N_2)$ as the gauge symmetry, the diffeomorphism group of $N_2$ \cite{yoon92}.  Let us introduce a coordinate system $\{u,v,y^a : a=2,3\}$ on $E_4$, where $\{u, v\}$ and $\{ y^a\} $ are coordinates on $M_{1+1}$ and  $N_2$, respectively. If we slice the spacetime $E_4$ by a family of null hypersurfaces and label
each of the null hypersurfaces by $u={\rm constant}$, then the most general line element is given by \cite{dinverno78}
\begin{eqnarray}
& & ds^2 =  0 \cdot dv^{2}  -2f dudv -2hdu^2+\phi_{ab}(dy^a+A_+^a du+A_-^a dv) \nonumber \\
& & \hspace{0.4in} \times (dy^b+A_+^b du+A_-^b dv).         \label{aborigin}
\end{eqnarray}
Let the subscripts $+$ and $-$ denote $u$ and $v$, respectively. Then, the horizontal lifts $\hat{\partial}_\pm$ 
of the tangent vector fields $\partial_\pm$ defined by
\begin{eqnarray}
\hat{\partial}_+=\partial_+ -A_+^a \partial_a, \quad
\hat{\partial}_-=\partial_- -A_-^a \partial_a                  \label{people}
\end{eqnarray}
are orthogonal to the vector field $\partial_a$ tangent to  $N_2$,
where $A_{\pm}^a$ are the connections valued in the Lie algebra of {\it diff}$(N_2)$.

The hypersurface $v={\rm constant}$ is ruled by $\hat{\partial}_+$
whose norm is $-2h$, which can be either positive, zero, or negative.
In this paper, the sign is chosen as
\be
-2h>0,                                              \label{near}
\ee
so that $v= {\rm constant}$ is a spacelike hypersurface. 
The hypersurface $u={\rm constant}$ is an out-going null hypersurface
ruled by the null vector field $\hat{\partial}_-$.
The intersection of two hypersurfaces $u,v={\rm constant}$ defines
a spacelike two-surface $N_{2}$ labeled by $y^{a}$ 
with a positive-definite metric $\phi_{ab}$ on it.

Let $n$ and $l$ be the in- and out-going null vector fields defined by
\begin{eqnarray}
n= k\Big( \hat{\partial}_+ - \frac{h}{f} \hat{\partial}_- \Big), \quad 
l= \frac{k}{f} \hat{\partial}_- ,
\end{eqnarray}
respectively, where $n$ and $l$ are normalized as
\begin{equation}
<n,l> \ = \ -k^2,   \quad k = {\rm constant} > 0 .                             \label{ournorm}
\end{equation} 
The positive-definite metric $\phi_{ab}$ on the two-surface $N_{2}$ can be written as a product of the area element $e^\sigma$ and the conformal 
two-metric $\rho_{ab}$, 
\begin{eqnarray}
\phi_{ab}=e^\sigma \rho_{ab}, \quad \det \rho_{ab}=1.
\end{eqnarray}
The {\it diff}($N_2$)-covariant derivative of a {\it diff}($N_2$)-tensor density ${T_{ab \cdots}}^{cd \cdots}$ with weight $w$ is defined by
\begin{eqnarray}
D_\pm {T_{ab \cdots}}^{cd \cdots} := \partial_\pm {T_{ab \cdots}}^{cd \cdots} -  {[A_\pm,T]_{ab \cdots}}^{cd \cdots},
\end{eqnarray}
where ${[A_\pm,T]_{Lab \cdots}}^{cd \cdots}$ is the Lie derivative of ${T_{ab \cdots}}^{cd \cdots}$ with respect to $A_\pm := A_\pm^a \partial_a$. The functions $h$ and $f$ are scalars, and $e^\sigma$ and $\rho_{ab}$ are densities with weight $1$ and $-1$ under the
{\it diff}($N_2$) transformations, respectively, so that one finds that
\begin{eqnarray}
& & D_\pm h =  \partial_\pm h -A_\pm^a \partial_a h 
= \hat{\partial}_{\pm}h,    \label{greenland}\\
& & D_\pm f = \partial_\pm f -A_\pm^a \partial_a f
=\hat{\partial}_{\pm}f,       \label{shark}\\
& & D_\pm \sigma = \partial_\pm \sigma - A_\pm^a \partial_a \sigma 
 -\partial_a A_\pm^a ,                                          \label{cod}\\
& & D_\pm \rho_{ab} = \partial_\pm \rho_{ab} -A_\pm^c \partial_c \rho_{ab}-\rho_{cb}\partial_a A_\pm^c  - \rho_{ac}\partial_b A_\pm^c +(\partial_c A_\pm^c)\rho_{ab}.  \label{creel}
\end{eqnarray}

The Einstein-Hilbert action is given by the spacetime 
integral of the following Lagrangian density, up to the total divergence,
\begin{eqnarray}
& & \hspace{-0.6in} 
\mathcal{L} = \frac{e^\sigma}{f} \{-2(D_{-} \sigma)(D_- h)+(D_{+} \sigma)(D_- f)+(D_{-} \sigma)(D_+ f)\}+f e^\sigma R^{(2)} \nonumber\\
& & \hspace{-0.4in} 
+ e^\sigma \{(D_+ \sigma)(D_- \sigma)-\frac{1}{2} \rho^{ab} \rho^{cd} (D_+ \rho_{ac})(D_- \rho_{bd})\}-\frac{1}{2f} e^{2\sigma} \rho_{ab} F_{+-}^a F_{+-}^b\nonumber\\
& & \hspace{-0.4in} 
+\frac{h}{f}e^\sigma\{ -(D_- \sigma)^2 + \frac{1}{2} \rho^{ab} \rho^{cd} (D_- \rho_{ac})(D_- \rho_{bd})\} -\frac{1}{2f} \rho^{ab}(\partial_a f)(\partial_b f). \label{lag}
\end{eqnarray}
Here $R^{(2)}$ is the Ricci scalar of $N_2$, and $F_{+-}^a$ is defined by
\bea
F_{+-}^a = \partial_+ A_-^a - \partial_- A_+^a - [A_+, A_-]_{\rm L}^a,                  \label{F+-}
\eea 
where 
\bea
[A_+, A_-]_{\rm L}^a = A_{+}^{b} \partial_{b} A_{-}^{a} - A_{-}^{b} \partial_{b} A_{+}^{a}. 
\eea

In order to write down the Einstein's equations in the first-order form, we need to identify the time coordinate. The natural time in this formalism is the {\it retarded time} $v$ \cite{dirac1949}, and the metric on the hypersurface $\Sigma$ defined by
$v=$ constant is given by
\begin{equation}
ds^2|_{\Sigma}= -2h du^2+e^{\sigma} \rho_{ab}(dy^a+A_+^a du ) (dy^b+A_+^b du).         \label{hyper}
\end{equation}
Thus, the configuration variables on $\Sigma$ are $q^I = (h, \sigma ,A_+^a, \rho_{ab})$, and functions ``0'', $f$, and $A_-^a$ in the metric (\ref{aborigin}) 
are the Lagrange multipliers. The momenta 
$\pi_{I}=(\pi_{h}, \pi_\sigma, \pi_a, \pi^{ab})$ conjugate to $q^I$ are found to be
\begin{eqnarray}
&\pi_h = {\delta \mathcal{L}  \over \delta (\partial _{-} h ) } =  -\frac{2}{f} e^\sigma D_- \sigma, \label{pih}  \\
&\pi_\sigma = {\delta \mathcal{L}  \over \delta (\partial_{-} \sigma )} 
= -\frac{2}{f} e^\sigma D_- h-\frac{2h}{f}e^\sigma D_- \sigma+e^\sigma D_+ \sigma 
+ 2 e^\sigma D_+ \ln f,  \label{pisigma}  \\
&\pi_a = {\delta \mathcal{L}  \over \delta (\partial_{-} A_+^a)} = \frac{1}{f} e^{2\sigma}\rho_{ab} F_{+-}^b - e^\sigma \partial_a \ln f, \label{pia}  \\
&\pi^{ab}= {\delta \mathcal{L}  \over \delta (\partial_{-} \rho_{ab})} = \frac{h }{f} e^\sigma \rho^{ac} \rho^{bd} D_- \rho_{cd}-\frac{1}{2} e^\sigma \rho^{ac} \rho^{bd} D_+ \rho_{cd}.
\label{piab}
\end{eqnarray}
Notice that the conjugate momentum $\pi^{ab}$ is traceless,
\bea
\rho_{ab}\pi^{ab} = 0.             \label{trace}
\eea
The first-order form of the above action is found to be 
\begin{eqnarray}
S =\int dv du d^2 y &\{ 
\pi_h \partial_{-} h +  \pi_\sigma \partial_- \sigma + \pi_a \partial_- A_+^a +\pi^{ab} \partial_- \rho_{ab} \nonumber \\
& - 0   \cdot C_{+} - f \ C_{-}  - A_-^a C_a  \},
\label{action}
\end{eqnarray} 
where $C_+$,  $C_-$, and $C_a$  are the constraints given by
\begin{eqnarray}
& & \hspace{-0.4in}
C_+ := \pi^{ab}D_+ \rho_{ab} + \pi_\sigma D_+ \sigma -hD_+ \pi_h -\partial_+ (h\pi_h 
+2e^\sigma D_+ \sigma) \nonumber\\
& & + \partial_a (h\pi_h A_+^a +2A_+^a e^\sigma D_+ \sigma + 2he^{-\sigma}\rho^{ab}\pi_b +2\rho^{ab}\partial_b h ) =0, \label{cplus}\\
& & \hspace{-0.4in}
C_- := \mathcal{H}-\partial_+ \pi_h +\partial_a (A_+^a \pi_h +e^{-\sigma}\rho^{ab} \pi_b ) =0, \label{cminus}\\
& & \hspace{-0.4in}
C_a := -\partial_+ \pi_a + \partial_b (A_+^b \pi_a) +\pi_b \partial_a A_+^b + \pi_\sigma \partial_a \sigma - \partial_a \pi_\sigma \nonumber \\
& & +\pi_h \partial_a h +\pi^{bc} \partial_a \rho_{bc} - \partial_b (\pi^{bc} \rho_{ac})- \partial_c (\pi^{bc} \rho_{ab}) + \partial_a (\pi^{bc} \rho_{bc}) = 0,  \label{cam}
\end{eqnarray}
and the function $\mathcal{H}$ in (\ref{cminus}) is defined as
\bea
& & \mathcal{H} := -\frac{1}{2}e^{-\sigma} \pi_\sigma \pi_h +\frac{1}{4}he^{-\sigma} \pi_h^2 -\frac{1}{2}e^{-2\sigma} \rho^{ab}\pi_a \pi_b -e^\sigma R^{(2)}\nonumber \\
& & \hspace{+0.4in}
+\frac{1}{2h}e^{-\sigma}\rho_{ac} \rho_{bd} \pi^{ab} \pi^{cd} +\frac{1}{8h} e^\sigma \rho^{ab} \rho^{cd} (D_+ \rho_{ac})(D_+ \rho_{bd}) \nonumber\\
& & \hspace{+0.4in}
+ \frac{1}{2h}\pi^{ab} D_+ \rho_{ab} +\frac{1}{2}\pi_h D_+ \sigma .
\eea
The constraints (\ref{cplus}), (\ref{cminus}), and (\ref{cam}) are the first-class constraints \cite{yoon14}, and  Lagrange multipliers ``0'', $f$, and $A_-^a$  enforce these constraints vanishing, i.e.,
\begin{eqnarray}
C_{+} = C_{-} = C_{a} = 0,  \label{constraints}
\end{eqnarray}
respectively.

The evolution equations of the conjugate momenta $\pi_h$, $\pi_\sigma$, $\pi_a$, and $\pi^{ab}$ in the retarded time $v$ are found to be
\bea
& & \hspace{-0.4in}
D_{-} \pi_h =  -\frac{f}{4} e^{-\sigma}\pi_h^2 +\frac{f}{2h^2}e^{-\sigma}\rho_{ac}\rho_{bd}\pi^{ab}\pi^{cd} +\frac{f}{2h^2}\pi^{ab}D_+ \rho_{ab} \nonumber\\
& & + \frac{f}{8h^2}e^\sigma \rho^{ac} \rho^{bd} (D_+ \rho_{ab})  (D_+ \rho_{cd}), \label{fish}\\
& & \hspace{-0.4in}
D_{-} \pi_\sigma  =  -\frac{f}{2}e^{-\sigma} \pi_\sigma \pi_h +\frac{1}{4}fh e^{-\sigma} \pi_h^2 
- f e^{-2\sigma}\rho^{ab}\pi_a \pi_b + \frac{f}{2}D_+ \pi_h \nonumber \\
&&+ \frac{f}{2h}e^{-\sigma}\rho_{ac} \rho_{bd} \pi^{ab} \pi^{cd} 
- \frac{f}{8h}e^\sigma \rho^{ac} \rho^{bd} (D_+ \rho_{ab})(D_+ \rho_{cd}) \nonumber\\
&&+\frac{1}{2} \pi_h D_+ f - e^{-\sigma} \rho^{ab} \pi_a \partial_b f 
+ \partial_a (\rho^{ab} \partial_b f), \label{cookie}\\
& & 
\hspace{-0.4in}
D_{-}\pi_a = \frac{f}{2}\pi_h\partial_a \sigma -\frac{f}{2}\partial_a \pi_h 
+ \frac{1}{2} \pi_h \partial_a f                               \nonumber\\
&&+\frac{f}{2h}\{\pi^{bc} \partial_a \rho_{bc}
+ \frac{1}{2}e^\sigma \rho^{bd} \rho^{ce} 
(D_+ \rho_{de})(\partial_a \rho_{bc})\}             \nonumber \\
&&-\partial_b (\frac{f}{h}\pi^{bc}\rho_{ac} 
+ \frac{f}{2h}e^\sigma \rho^{bc}D_+ \rho_{ac}),               \label{chip}\\
& & \hspace{-0.4in}
D_{-} \pi^{ab} = -\frac{f}{2}e^{-2\sigma}\rho^{ac}\rho^{bd}\pi_c \pi_d +\frac{f}{4}e^{-2\sigma}\rho^{ab}\rho^{cd}\pi_c \pi_d -\frac{f}{h}e^{-\sigma}\rho_{cd}\pi^{ac}\pi^{bd} \nonumber\\
&& 
 +D_+(\frac{f}{4h}e^\sigma \rho^{ac} \rho^{bd}D_+\rho_{cd} +\frac{f}{2h}\pi^{ab}) 
+\frac{f}{4h}e^\sigma \rho^{ac} \rho^{bd} \rho^{ef}(D_+ \rho_{ce})(D_+ \rho_{df})  \nonumber\\
&& 
+\frac{1}{2} \rho^{ac} \rho^{bd} \{(\partial_c f)(\partial_d \sigma) 
+(\partial_d f)(\partial_c \sigma )\}
-\frac{1}{2} \rho^{ab} \rho^{cd} (\partial_c f)(\partial_d \sigma) \nonumber\\
& &
-\frac{1}{2} e^{-\sigma}\rho^{ac} \rho^{bd}  ( \pi_c \partial_d f + \pi_d \partial_c f )
+\frac{1}{2} e^{-\sigma}\rho^{ab} \rho^{cd} \pi_c \partial_d f .
\label{wasabi}
\eea
The 12 equations (\ref{pih}), $\cdots$, (\ref{piab}), and (\ref{fish}), $\cdots$, (\ref{wasabi}) 
are the first-order forms of the 6 second-order evolution equations 
in the $v$-time. Together with the 4 constraint equations (\ref{cplus}), (\ref{cminus}),
and (\ref{cam}) which contain at most first-order $v$-derivatives, 
they are equivalent to the vacuum Einstein's equations $R_{AB} = 0$~($A,B=+,-, a$).

The physical contents of the constraint equations can be best
exhibited by integrating them over an arbitrary closed two-surface $N_2$. If we multiply (\ref{cam}) with an arbitrary tangent vector field $\xi=\xi^a \partial_a$ on $N_2$, then it becomes
\bea
& & 
\xi^a C_a = \pi^{ab} \mathcal{L}_\xi \rho_{ab} + \pi_\sigma \mathcal{L}_\xi \sigma +\pi_h \mathcal{L}_\xi h +\pi_a \mathcal{L}_\xi A_+^a +\pi_a \partial_+ \xi^a \nonumber\\
&& \hspace{0.4in}
- \partial_+(\xi^a \pi_a) + \partial_a (-\xi^a \pi_\sigma - 2\pi^{ab}\xi^c \rho_{bc}+A_+^a \xi^b \pi_b)=0,                           \label{contracted}
\eea
where $\mathcal{L}_\xi$ is the Lie derivative along $\xi$. Notice that all the constraint equations 
\bea
C_+=C_-=\xi^a C_a=0                 \label{divergence}
\eea 
are partial differential equations of the divergence-type
\bea
\partial_{\mu} j^{\mu} -  \mathcal{F}  = 0 \quad (\mu=+, a).  \label{flux}
\eea
Integrating the divergence-type equation (\ref{flux}) over $N_2$ , we obtain
the balance equations of the form
\bea
\frac{\partial}{\partial u} Q(u,v) = \oint d^2 y \, \mathcal{F}  (u,v,y^{a}),   \label{clean}
\eea
where $Q(u,v)$ is the quasilocal charge defined by
\be
Q(u,v):=\oint d^2 y \ j^{+} (u,v,y^{a}),  
\ee
and $ \mathcal{F} (u,v, y^{a})$ is the corresponding flux.
Thus, if we integrate each of the constraint equations (\ref{cplus}), (\ref{cminus}), and
(\ref{contracted}) over $N_2$, then we obtain the following integro-differential equations:
\bea
& & 
\frac{\partial}{\partial u} U(u,v) = \frac{1}{16\pi} \oint  d^2 y 
( \pi^{ab}D_+ \rho_{ab} + \pi_\sigma D_+ \sigma -hD_+ \pi_h), \label{balance1}\\
& &
\frac{\partial}{\partial u} P(u,v) = \frac{1}{16\pi} \oint d^2 y \ \mathcal{H},  \label{balance2}\\
& & 
\frac{\partial}{\partial u} L(u,v;\xi) = \frac{1}{16\pi} \oint d^2 y (\pi^{ab} \mathcal{L}_\xi \rho_{ab} + \pi_\sigma \mathcal{L}_\xi \sigma  +\pi_h \mathcal{L}_\xi h \nonumber\\
&& \hspace{1in}
 +\pi_a \mathcal{L}_\xi A_+^a +\pi_a \partial_+ \xi^a),\label{balance3}
\eea
where $U(u,v)$, $P(u,v)$ and $L(u,v;\xi)$ are quasilocal quantities defined by
\bea
& & U(u,v) :=  \frac{1}{16\pi} \oint d^2 y (h\pi_h+2e^\sigma D_+ \sigma) +U_{0},\label{up} \\
& & P(u,v) := \frac{1}{16\pi} \oint d^2 y (\pi_h)+P_{0} ,\label{down} \\
& & L(u,v;\xi ) :=   \frac{1}{16\pi} \oint d^2 y (\xi^a \pi_a ) + L_{0} \label{strange},
\eea
and $U_{0}$, $P_{0}$ and $L_{0}$ are {\it undetermined integration constants}. 
The equations (\ref{balance1}), (\ref{balance2}), and (\ref{balance3}) are the balance equations that relate the rates  of changes in $U$, $P$, and $L$ to the corresponding fluxes as $u$ changes \cite{yoon04,ashtekar02}.

The quasilocal quantity $L(u,v;\xi )$ given by (\ref{strange})
can be expressed geometrically
as follows. Let us notice that the Lie bracket of the in- and out-going null vector fields $n$ and $l$ is given by
\bea
& & \hspace{-0.4in}
[n,l]_{\rm L} = [k\Big( \hat{\partial}_+ - \frac{h}{f} \hat{\partial}_- \Big) , \frac{k}{f} \hat{\partial}_-]_{\rm L} \nonumber\\
& & = k^2 [\hat{\partial}_+ , \frac{1}{f}\hat{\partial}_-]_{\rm L} - k^2 [ \frac{h}{f} \hat{\partial}_- , \frac{1}{f}\hat{\partial}_-]_{\rm L} \nonumber\\
& & = \frac{k^2}{f} [\hat{\partial}_+,\hat{\partial}_-]_{\rm L} 
+ k^2 (\hat{\partial}_+ \frac{1}{f}) \hat{\partial}_-
-\frac{k^2 h}{f} (\hat{\partial}_- \frac{1}{f} ) \hat{\partial}_- +\frac{k^2}{f} 
(\hat{\partial}_- \frac{h}{f}) \hat{\partial}_- \nonumber \\
& & = \frac{k^2}{f}[\partial_+ -A_+^a \partial_a ,\  \partial_- - A_-^b \partial_b]_{\rm L} 
- \frac{k^2}{f^2}(\hat{\partial}_+ f - \hat{\partial}_- h) \hat{\partial}_- \nonumber\\
& & = -\frac{k^{2}}{f}F_{+-}^a \partial_a - \frac{k}{f}(D_+ f - D_- h) l.  \label{news}
\eea
In the last line of the above equation, we used the equations
(\ref{greenland}), (\ref{shark}), and (\ref{F+-}).
Thus, the transverse component of
$[n,l]_{\rm L}$ is given by
\begin{eqnarray}
& & [n,l]_{\rm L}^{a} = -\frac{k^{2}}{f}F_{+-}^a  \nonumber\\
& & \hspace{0.4in}
 = - k^{2} (e^{-2\sigma} \rho^{ab}\pi_{b} 
+ e^{-\sigma} \rho^{ab}\partial_{b} \ln f ), \label{transnl}
\end{eqnarray}
where we used the equation (\ref{pia}).
Conversely, $\pi_{a} $ can be written as
\begin{equation}
\pi_{a} = -{1\over k^{2}}e^{2\sigma} \rho_{ab} [n,l]_{\rm L}^{b} 
- e^{\sigma} \partial_{a}\ln f .  \label{cafe}
\end{equation}
If we use the normalization (\ref{ournorm}) and the identity
$[n,l]_{{\rm L} a} = e^{\sigma} \rho_{ab} [n,l]_{\rm L}^{b}$, $\pi_{a}$
can be written in an invariant form
\begin{equation}
\pi_{a} =e^{\sigma} { [n,l]_{La} \over <n, l>}
- e^{\sigma} \tilde{\nabla}_{a}\ln f ,  \label{piecafe}
\end{equation}
where $\tilde{\nabla}_a$ is the covariant derivative on $N_2$.
Let $d\mu$ be  the invariant measure on $N_{2}$ given by
\be
d\mu =  d^2 y \ e^{\sigma}.
\ee
Then, one finds that 
\begin{eqnarray}
& & \oint d^2 y \ \xi^a \pi_a   
=  \oint d \mu \Big( \frac{< \xi , [n,l]_{\rm L} >}{<n,l>} 
- \xi^a \tilde{\nabla}_a  \ln{f} \Big)  \nonumber\\
&  &  \hspace{0.8in}
= \oint d\mu  \Big( \frac{< \xi , [n,l]_{\rm L} >}{<n,l>} 
 +   \ln{f} \tilde{\nabla}_a \xi^a \Big),  \label{simple}
\end{eqnarray} 
so that the equation (\ref{strange}) becomes
\begin{equation}
L(u,v;\xi ) 
= \frac{1}{16\pi} \oint d \mu \Big( \frac{< \xi , [n,l]_{\rm L} >}{<n,l>} 
+ \ln{f} \tilde{\nabla}_{a} \xi^a    \Big) 
+ L_{0} .                      \label{compact}
\end{equation}

%%%%%%%%%%%%%%%%%%%%%%%%%%%%%%%%%%%%%%%%%%%%%%%%%%%%%%%%%%%%%%%%%%%%%%%%%%%%%%%%%%%%%%%%%%

\section{Gauge-independence of quasilocal angular momentum} \label{gauge}

A two-surface $N_2$ in a 4-dimensional spacetime $E_4$ can be described as a level surface
$\{p \in E_4 \, | \, \lambda(p) = \nu (p) = 0 \}$, where $\lambda$ and $\nu$ are scalar functions defined on $E_4$. Since the null vector fields $n$ and $l$ are normal to $N_2$, they can be written locally as 
\bea
n_A = \nabla_A \lambda , \quad l_A = \nabla_A \nu ,
\eea
where $A= (+, -, a)$. The functions $\lambda$ and $\nu$, however, are not unique, and the quasilocal angular momentum must be independent of 
labeling two-surfaces $N_2$ in $E_4$. Let us consider two functions $\lambda'$ and $\nu'$ related to $\lambda$ and $\nu$ by
\bea
\lambda' = \alpha \lambda, \quad \nu' = \beta \nu,
\eea
where $\alpha$ and $\beta$ are arbitrary functions, so that
the level surface defined by $\{p \in E_4 \, | \, \lambda'(p) = \nu' (p) = 0 \}$ describes the same surface $N_2$ as before.  We will show that the necessary and sufficient condition that the quasilocal angular momentum (\ref{compact}) is independent of the choice of the arbitrary functions $\alpha$ and $\beta$ is 
\be
\tilde{\nabla}_{a}\xi^{a} = 0, 
\ee
where $\tilde{\nabla}_{a}$ is the covariant derivative on $N_{2}$. 
The null vector fields normal to $N_2$ can be also written as
\bea
n'_A = \nabla_A \lambda' |_{\lambda'=0} 
= (\alpha \nabla_A \lambda +\lambda \nabla_A \alpha ) |_{\lambda=0} = \alpha \nabla_A \lambda,   \nonumber\\
l'_A = \nabla_A \nu' |_{\nu'=0} 
= ( \beta \nabla_A \nu +\nu \nabla_A \beta )|_{\nu=0} 
= \beta \nabla_A \nu . 
\eea
In order to preserve the normalization condition $<n',l'>=-k^2$, $\beta$ must be related to $\alpha$ by
\bea
\beta = \frac{1}{\alpha}.                        \label{ab}
\eea
The covariant derivative $\tilde{\nabla}_a$ on $N_2$ is the projection of the covariant derivative $\nabla_A$ on the whole spacetime $E_4$, 
\bea
 \tilde{\nabla}_a = {h_a}^A \nabla_A, 
\eea
where ${h_a}^A$ is the projector from the tangent space of $E_4$ to the tangent space of $N_2$. Then, we have 
\bea
& & \hspace{-0.55in}
[n',l']_{La} = h_{aA}[n',l']_{\rm L}^A \nonumber \\
& & = h_{aA} (n'^B \nabla_B l'^A - l'^B \nabla_B n'^A) \nonumber\\
& & = h_{aA} \{ \nabla^B (\alpha \lambda) \nabla_B \nabla^A (\beta \nu) -\nabla^B (\beta \nu) \nabla_B \nabla^A (\alpha \lambda) \} |_{\lambda=\nu=0} \nonumber\\
& & = h_{aA} \{ \alpha (\nabla^B \lambda) \nabla^A \nabla_B (\beta \nu) -\beta (\nabla^B \nu) \nabla^A \nabla_B (\alpha \lambda) \} |_{\lambda=\nu=0} \nonumber\\
& & = h_{aA} \{ \alpha (\nabla^B \lambda) \nabla^A( \beta \nabla_B \nu + \nu \nabla_B \beta)                          \nonumber\\
& & \hspace{0.2in}
-\beta (\nabla^B \nu) \nabla^A 
   ( \alpha \nabla_B \lambda + \lambda \nabla_B \alpha) \} |_{\lambda=\nu=0}.
\label{twist1}
\eea
Notice that we have
\bea
\hspace{-0.2in}
h_{aA} (\nabla^B \lambda) \nabla^A (\nu \nabla_B \beta) 
= h_{aA} (\nabla^B \lambda) \{ (\nabla^A \nu) \nabla_B \beta + \nu \nabla^A \nabla_B \beta\}
=0,
\label{twist2}
\eea
because $\nu=0$ and $h_{aA} \nabla^A \nu = h_{aA} l^{A} = 0$ on $N_2$.  Likewise, we have
\bea
h_{aA} (\nabla^B \nu) \nabla^A (\lambda \nabla_B \alpha) = 0. 
\label{twist3}
\eea
If we use (\ref{twist2}) and (\ref{twist3}), then (\ref{twist1}) becomes
\bea
[n',l']_{La} &= h_{aA} \{ \alpha (\nabla^B \lambda) (\nabla^A \beta \cdot \nabla_B \nu + \beta \nabla^A \nabla_B \nu)|_{\lambda=\nu=0} \nonumber\\
& \ \ \ -\beta (\nabla^B \nu)  (\nabla^A \alpha \cdot \nabla_B \lambda + \alpha \nabla^A \nabla_B \lambda) \} |_{\lambda=\nu=0}.
\label{twist4}
\eea
This equation becomes, by the relation (\ref{ab}),
\bea
& & 
[n',l']_{La} = h_{aA} ( \nabla^B \lambda \cdot \nabla^A \nabla_B \nu - \nabla^B \nu \cdot \nabla^A \nabla_B \lambda )|_{\lambda=\nu=0} \nonumber\\
& &  \hspace{0.8in}
-2h_{aA} (\nabla^A \ln{\alpha}) (\nabla^B \lambda \cdot \nabla_B \nu)|_{\lambda=\nu=0} \nonumber\\
& & \hspace{0.6in} = h_{aA} \{ \nabla^A (\nabla^B \lambda \cdot \nabla_B \nu ) -2 (\nabla^A \nabla^B \lambda) \nabla_B \nu \}|_{\lambda=\nu=0} \nonumber\\
& & \hspace{0.8in} 
-2h_{aA} (\nabla^A \ln{\alpha}) (\nabla^B \lambda \cdot \nabla_B \nu)|_{\lambda=\nu=0}.
\eea
By the identity $\nabla^B \lambda \cdot \nabla_B \nu = n^B l_B = -k^2$, the above equation becomes
\bea
[n', l']_{La} = -2 (\tilde{\nabla}_a \nabla^B \lambda ) \nabla_B \nu|_{\lambda=\nu=0} 
+ 2 k^2 \tilde{\nabla}_a \ln \alpha. \label{comm}
\eea
When  $\alpha =1$, $n'$ and $l'$ become $n$ and $l$, respectively, and the equation (\ref{comm}) becomes
\bea
[n, l]_{La} = -2 (\tilde{\nabla}_a \nabla^B \lambda ) \nabla_B \nu|_{\lambda=\nu=0}. \label{newcomm}
\eea
Thus, the equation  (\ref{comm}) can be written as
\bea
[n', l']_{La} = [n,l]_{La} + 2 k^2 \tilde{\nabla}_a \ln \alpha,   \label{new}
\eea
so that we have
\bea
\oint d \mu  \frac{<\xi,[n',l']_{\rm L}>}{<n',l'>}  
& = &
\oint d \mu  \Big( \frac{<\xi,[n,l]_{\rm L}>}{<n,l>} -2 \xi^{a}\tilde{\nabla}_{a}\ln \alpha \Big) \nonumber\\
& =& \oint d  \mu \Big(\frac{<\xi,[n,l]_{\rm L}> }{<n,l>} 
 +  2 \ln \alpha  \tilde{\nabla}_a \xi^a \Big) .   \label{pork}
\eea
Then, $L'(u,v;\xi )$ becomes
\bea
L'(u,v;\xi )
& = & 
{1\over 16\pi} \oint d  \mu \Big\{ 
\frac{<\xi,[n,l]_{\rm L}> }{<n,l>} 
 +  \ln ( f \alpha^{2} )  \tilde{\nabla}_a \xi^a \Big\}
+ L_{0}   \nonumber\\
& =& 
L (u,v;\xi ) + {1\over 16\pi} \oint d  \mu \ 
\ln ( \alpha^{2} )  \tilde{\nabla}_a \xi^a. \label{lobo}
\eea 
Thus, we find that the quasilocal angular momentum 
is insensitive to the ways of labeling a given two-surface in the spacetime, i.e., 
\bea
L'(u,v; \xi) = L(u,v;\xi),
\eea
if and only if 
\bea
\tilde{\nabla}_a \xi^a = 0.              \label{suff}
\eea
Therefore, the quasilocal angular momentum is given by
\begin{equation}
L(u,v;\xi ) 
= \frac{1}{16\pi} \oint d \mu  \frac{ < \xi,  [n,l]_{\rm L} > }{<n,l>}  
+ L_{0} \quad  (  \tilde{\nabla}_a \xi^a = 0),            \label{ensenble}
\end{equation}
where $L_{0}$ is an undetermined integration constant. 

It is worth mentioning that the condition (\ref{suff}) is equivalent to
the condition that ensures the gauge-independence of quasilocal angular momentum defined by C.M. Liu and S.T. Yau \cite{liu_yau03}.
Notice that the condition (\ref{suff}) is trivially satisfied when $\xi$ is a Killing vector field on $N_{2}$, 
\bea
\tilde{\nabla}_{(a} \xi_{b)} = 0.   \label{killing}
\eea
Obviously, the converse is not true; the divergence-free vector field  $\xi$ 
need not be a Killing vector field of $N_{2}$. Thus, the quasilocal angular momentum  $L(u,v;\xi)$ is well-defined even when $N_{2}$ does not admit any isometries, provided that $\xi$ is divergence-free!

One might ask whether it is always possible to foliate the whole spacetime $E_{4}$ by
a given 2-dimensional surface $N_{2}$ and question the stability of such foliation.
Works related to these issues can be found in \cite{gittel2013,chrusciel2013,huang2010}, where a ``rigid-sphere foliation\rq\rq{} with a non-constant mean curvature in a 
3-dimensional Riemannian space was discussed. 
It is also possible to refine the quasilocal angular momentum $L(u,v;\xi)$ by further restricting divergence-free vector field $\xi^{a}$ on $N_{2}$, which can be represented as
\be
\xi^{a} = \epsilon^{ab} \tilde{\nabla}_{b} f + \xi_{0}^{a}
\ee
where $\xi_{0}^{a}$ is the harmonic vector field. By selecting a suitable function $f$ and 
$\xi_{0}^{a}$ on $N_{2}$,
one may reduce the quasilocal angular momentum $L(\xi)$ to a proper subclass that has special properties\cite{gittel2013}.

\begin{figure}[!bt]
  \centering
      \includegraphics[width=0.5\textwidth]{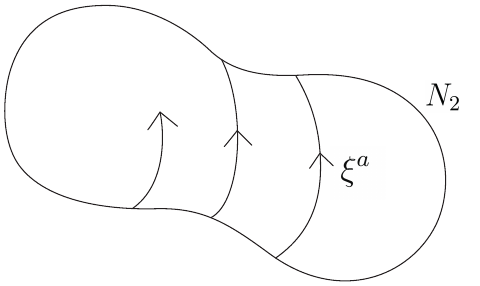}
  \caption{A divergence-free vector field $\xi^{a}$ need not be a Killing vector field.}
\end{figure}

%%%%%%%%%%%%%%%%%%%%%%%%%%%%%%%%%%%%%%%%%%%%%%%%%%%%%%%%%%%%%%%%%%%%%%%%%%%%%%%%%%%%%%%

\section{Geometric interpretations }\label{geometric}

The quasilocal quantities defined in (\ref{up}) and (\ref{down}) can be interpreted geometrically also. 
The equation (\ref{up}) becomes, by the equation (\ref{pih}),
\bea
& & \hspace{-0.5in}
U(u,v) = \frac{1}{8\pi} \oint d^2 y e^\sigma (D_+ \sigma - \frac{h}{f} D_- \sigma) +U_{0} \nonumber \\
& & = \frac{1}{8\pi k} \oint d^2 y \ \mathcal{L}_n e^\sigma +U_{0} \nonumber\\
& &  = \frac{\mathcal{L}_n \mathcal{A}}{8\pi \sqrt{-<n,l>}} +U_{0} ,             \label{ua}
\eea
where $\mathcal{A}$ is the area of $N_{2}$, 
\bea
\mathcal{A} = \oint d\mu = \oint d^2 y \ e^\sigma ,       \label{area}
\eea 
and we used the fact that the Lie derivative $\mathcal{L}_n$ along the null normal $n$ 
and $d^2 y$ integration are interchangeable in the last line.
The equation (\ref{down}) becomes
\bea
& & 
P(u,v) =  -\frac{1}{8\pi} \oint d^2 y  \frac{e^\sigma}{f} D_- \sigma +P_{0} \nonumber\\
& & \hspace{0.5in} 
= -\frac{1}{8\pi k} \oint d^2 y \mathcal{L}_l e^\sigma +P_{0}\nonumber\\
& & \hspace{0.5in} 
=-\frac{\mathcal{L}_l \mathcal{A}}{8\pi \sqrt{-<n,l>}}+P_{0}.            \label{pa}
\eea

In order to evaluate these quasilocal quantities relative to a background spacetime, the integration constants $U_{0}$ and $P_{0}$ must be chosen
appropriately depending on the reference spacetime. 
There are several works on how to fix these constants when the background spacetime is the Minkowski spacetime $\mathbb{R}^{3,1}$. 
With the Minkowski spacetime  as the background spacetime,
we will show that our quasilocal quantities reduce to the well-known expressions of quasilocal energy and momentum, provided that the integration constants are chosen 
appropriately \cite{brown_york92,brown_york93,liu_yau03,wang_yau09,chen_wang_yau13}.

Let $e_{0}$ be a timelike unit vector field normal to $\Sigma \subset E_{4}$, and 
$e_{1}$ a spacelike unit outward normal vector field on $N_{2}\subset \Sigma$, respectively. They are related to the null vector fields $n$ and $l$,
\be
e_{0}={1\over \sqrt{2}k} (l + n),  \quad
e_{1}={1\over \sqrt{2}k} (l - n).       \label{unit}
\ee
Then, from (\ref{ua}) and (\ref{pa}), one finds that
\bea
& & 
E :={1\over \sqrt{2}} ( U + P ) = -{1\over 8 \pi} \oint  d \mu \ {\rm tr}\, k
+ {1\over \sqrt{2}}(U_{0}+ P_{0}),   \label{bya}\\
& & 
Q:={1\over \sqrt{2}} ( U - P ) = {1\over 8 \pi} \oint  d \mu \ {\rm tr}\, p
+ {1\over \sqrt{2}}(U_{0}- P_{0}),   \label{bya}
\eea
where ${\rm tr} \, k$ and ${\rm tr} \, p$ are the traces of the extrinsic curvature of $N_{2}$ with respect to $e_{1}$ and  $e_{0}$, respectively.

When  $N_{2}$ is a topological two-sphere such that it can be isometrically embedded in $\mathbb{R}^{3}\subset \mathbb{R}^{3,1}$ by the Weyl embedding theorem, one can define a timelike unit vector field 
$\bar{e}_{0}$ normal to $\mathbb{R}^{3} \subset \mathbb{R}^{3,1}$ and 
a spacelike unit outward normal vector field 
$\bar{e}_{1}$ on  $N_{2}\subset \mathbb{R}^{3}$. 
Let  ${\rm tr} \, \bar{k}$ and ${\rm tr} \, \bar{p}$ be the trace of the extrinsic curvature of $N_{2} \subset\mathbb{R}^{3}$ with respect to $\bar{e}_{1}$ and $\bar{e}_{0}$, respectively. By construction, it follows trivially that ${\rm tr} \, \bar{p}=0$.

If the integration constant $E_{0}$ is chosen such that
\be
E_{0}:= {1\over \sqrt{2}}(U_{0}+ P_{0}) = {1\over 8 \pi} \oint  d \mu \ {\rm tr}\, \bar{k},
\ee
then one finds that
\be
E = -{1\over 8 \pi} \oint  d \mu \  
( {\rm tr}\, k -{\rm tr}\, \bar{k} ),
\ee
which is precisely the quasilocal energy of Brown-York \cite{brown_york92,brown_york93,liu_yau03,wang_yau09,chen_wang_yau13}. 
If the integration constant $Q_{0}$ is chosen as
\be
Q_{0}:= {1\over \sqrt{2}}(U_{0} - P_{0}) = {1\over 8 \pi} \oint  d \mu \ {\rm tr}\, \bar{p}  =  0,
\ee
then it follows that
\be
Q = {1\over 8 \pi} \oint  d \mu \  
 {\rm tr}\, p,
\ee
which is the quasilocal momentum of Brown-York. 
Detailed discussions on the existence of isometric embedding of $N_{2}$ into the Minkowski spacetime $\mathbb{R}^{3,1}$ and properties such as the invariance and positivity of quasilocal mass relative to the background Minkowski spacetime can be found in \cite{liu_yau03,wang_yau09,chen_wang_yau13}.

%%%%%%%%%%%%%%%%%%%%%%%%%%%%%%%%%%%%%%%%%%%%%%%%%%%%%%%%%%%%%%%%%%%%%%%%%%%%%%%%%%%%%%%%

\section{Poisson algebra of quasilocal angular momentum}\label{poisson}

As was discussed in the previous section, the two-surface integral $L(u,v;\xi)$  depends on the divergence-free vector field $\xi$. Let us assume that the integration constant $L_{0}=0$. By the constraint equation (\ref{balance3}), 
the two-surface integral can be also written as a 3-dimensional integral of the associated flux, namely, 
\bea
L(u,v;\xi )&=& \frac{1}{16\pi} \int du \oint d^2 y (\pi^{ab} \mathcal{L}_{\xi} \rho_{ab} + \pi_\sigma \mathcal{L}_{\xi} \sigma  \nonumber\\
&&\hspace{0.5in} +\pi_h \mathcal{L}_{\xi} h+\pi_a \mathcal{L}_{\xi} A_+^a 
+\pi_a \partial_+ {\xi}^a)
\quad (\tilde{\nabla}_{a}\xi^{a} =0). 
\label{QAM}
\eea
We prove the following theorem. \\

\noindent {\bf Theorem.} Let $\xi$ and $\eta$ be two arbitrary vector fields
tangent to $N_{2}$. 
The Poisson bracket of quasilocal angular momentums associated with $\xi$ and $\eta$ is 
given by the quasilocal angular momentum associated with the Lie bracket of $\xi$ and $\eta$ up to a constant factor. Precisely, it is given by
\bea
\{ L(\xi), L(\eta) \}_{\rm {P.B.}} = \frac{1}{16\pi} L([\xi, \eta]_{\rm L}),    \label{water}
\eea
where the Poisson bracket of any two functionals  $A$ and $B$ on a 
$v=$ constant hypersurface $\Sigma$ is defined by
\bea
\{A,B\}_{\rm P.B.} = \int du \oint d^2 y \left( \frac{\delta A}{\delta q^I}\frac{\delta B}{\delta p_I}
-\frac{\delta B}{\delta q^I}\frac{\delta A}{\delta p_I} \right).
\eea 
\\
Proof. 
Let us notice that, for any {\it diff}$(N_2)$-tensor density $T_{b \cdots}^{\ c \cdots} $ with weight 1, the following identity 
\bea
\mathcal{L}_\xi T_{ b \cdots}^{\ c \cdots} = \xi^a \partial_a T_{ b \cdots}^{\ c \cdots} 
 + T_{ b \cdots}^{\ c \cdots} \partial_a \xi^a = \partial_a (\xi^a T_{ b \cdots}^{\ c \cdots}) \label{tea}
\label{id1}
\eea
holds. Using this identity, one finds, after straightforward calculations,
\bea
\hspace{-0.6in}
\{L(\xi), L(\eta) \}_{\rm P.B.} 
& & =  - \Big(\frac{1}{16\pi} \Big)^2 \int du \oint d^2 y \Big\{ (\mathcal{L}_{\xi} \pi^{ab}
)(\mathcal{L}_{\eta} \rho_{ab}) + (\mathcal{L}_{\xi} \pi_{\sigma}
)(\mathcal{L}_{\eta} \sigma) \nonumber\\
& &  + (\mathcal{L}_{\xi} \pi_{h})(\mathcal{L}_{\eta} h) 
+(\mathcal{L}_{\xi} \pi_{a})(\mathcal{L}_{\eta} A_+^{a} + \partial_+ \eta^a) - (\xi \leftrightarrow \eta) \  \Big\} \nonumber\\
& & \hspace{-0.6in} 
= \Big(\frac{1}{16\pi} \Big)^2 \int du \oint d^2 y \Big\{ (\pi^{ab}
\mathcal{L}_{\xi} \mathcal{L}_{\eta} \rho_{ab} 
+ \pi_{\sigma} \mathcal{L}_{\xi} \mathcal{L}_{\eta} \sigma 
+ \pi_{h} \mathcal{L}_{\xi} \mathcal{L}_{\eta} h   \nonumber\\
& & 
+\pi_{a} \mathcal{L}_{\xi}\mathcal{L}_{\eta} A_+^{a} 
+\pi_a \mathcal{L}_\xi \partial_+ \eta^a 
-   (\xi \leftrightarrow \eta) \Big\}  \nonumber\\
& & \hspace{-0.6in} 
= \Big(\frac{1}{16\pi} \Big)^2 \int du \oint d^2 y \Big\{ (\pi^{ab}
\mathcal{L}_{[\xi, \eta ]_{\rm L}} \rho_{ab} 
+ \pi_{\sigma} \mathcal{L}_{[\xi, \eta ]_{\rm L}} \sigma   
+ \pi_{h} \mathcal{L}_{[\xi, \eta ]_{\rm L}} h \nonumber\\
& &
+\pi_{a} \mathcal{L}_{[\xi, \eta ]_{\rm L}} A_+^{a} 
+ \pi_a \partial_+ [\xi, \eta]_{\rm L}^a \Big\} \nonumber\\
& & \hspace{-0.6in} 
=  \frac{1}{16\pi} L([\xi, \eta ]_{\rm L}).   \label{cup}
\eea
In the second to the last line of the above equation, we used the following identity
\bea
\hspace{-0.4in}
\mathcal{L}_\xi \partial_+ \eta^a - \mathcal{L}_\eta \partial_+ \xi^a 
&&= \xi^b \partial_b \partial_+ \eta^a - (\partial_+ \eta^b) \partial_b \xi^a - (\xi \leftrightarrow \eta) \nonumber\\
&&= \partial_+ (\xi^b \partial_b \eta^a ) - (\partial_+ \xi^b)(\partial_b \eta^a)
- (\partial_+ \eta^b) \partial_b \xi^a - (\xi \leftrightarrow \eta) \nonumber \\
&&= \partial_+ (\xi^b \partial_b \eta^a - \eta^b \partial_b \xi^a) \nonumber\\
&&= \partial_+ [\xi, \eta]_{\rm L}^a.                                     \label{id2}
\eea
Let us show that the Lie bracket of any two divergence-free 
vector fields is also divergence-free,
\be
\tilde{\nabla}_{a} [ \xi , \eta]_{L}^{\ a} =0 \ \ {\rm if} \ \ 
\tilde{\nabla}_{a}\xi^{a} = \tilde{\nabla}_{a}\eta^{a} =0.  \label{dog}
\ee
Notice that the covariant divergence of the Lie bracket 
$[ \xi , \eta ]_{L}^{\ a}$ defined by
\be
[ \xi , \eta ]_{L}^{\ a} = \xi^{b} \tilde{\nabla}_{b} \eta^{a} 
- \eta^{b} \tilde{\nabla}_{b} \xi^{a}
\ee
is given by
\be
\tilde{\nabla}_{a} [ \xi , \eta ]_{L}^{\ a}
= \xi^{b}\tilde{\nabla}_{a} \tilde{\nabla}_{b} \eta^{a} 
- \eta^{b}\tilde{\nabla}_{a} \tilde{\nabla}_{b} \xi^{a} .
\ee
Using the 2-dimensional curvature tensor $R^{(2)\ c}_{abd}$ and the Ricci tensor 
$R^{(2)}_{ad}$ defined as
\be
\tilde{\nabla}_{[ a}\tilde{\nabla}_{b]} \eta^{c} = R^{(2)\ c}_{abd} \eta^{d}, \quad 
R^{(2)\ b}_{abd} =  R^{(2)\ }_{ad},
\ee
respectively, one finds that
\bea
& & 
\xi^{b} \tilde{\nabla}_{a}\tilde{\nabla}_{b}\eta^{a} 
=\xi^{b} \tilde{\nabla}_{[a}\tilde{\nabla}_{b]}\eta^{a} 
=-R_{ab}^{(2)}\ \xi^{a}\eta^{b}, \nonumber\\
& & 
\eta^{b} \tilde{\nabla}_{a}\tilde{\nabla}_{b}\xi^{a} 
=-R_{ab}^{(2)}\ \xi^{a} \eta^{b}, 
\eea
for divergence-free vector fields $\xi^{a}$ and $\eta^{a}$.
Therefore, it follows that
\be
\tilde{\nabla}_{a} [ \xi , \eta ]_{L}^{\ a} = 0.
\ee
It follows that, given two gauge-independent quasilocal angular momenta $L(\xi)$ and $L(\eta)$, their Poisson bracket $L([\xi, \eta ]_{\rm L})$ is also gauge-independent.

%%%%%%%%%%%%%%%%%%%%%%%%%%%%%%%%%%%%%%%%%%%%%%%%%%%%%%%%%%%%%%%%%%%%%%%

\section{Invariant quasilocal angular momentum $L^{2}$}\label{invariant}

The quasilocal angular momentum $L(\xi )$ given by (\ref{strange}) or (\ref{compact}) depends on the divergence-free vector field $\xi^{a}$ defined on $N_{2}$. 
By the Poincar{\'{e}}-Hopf theorem, for any compact regular 2-dimensional manifold with non-zero Euler characteristic, any continuous tangent vector field has at least one zero.
Therefore, $L(\xi )$ fails to ``measure'' the rotation of gravitational fields
at these singular points of $\xi^{a}$, as was discussed in Introduction.

In this section, we propose a definition of {\it invariant} quasilocal angular momentum $L^{2}$ of gravitational fields, which is free from the singularity of the divergence-free tangent vector field $\xi^{a}$ and commutes with $L(\xi)$, 
\be
 \{ L^{2}, L(\xi)  \}_{\rm P.B.}  = 0.  \label{mouse}
\ee 
It is defined as an invariant integral
over $N_{2}$,
\begin{equation}
L^2 (u,v) := \frac{1}{96 \pi} \Big(\frac{\mathcal{A}}{4\pi} \Big)^{2} \oint d^{2} y \ 
e^{-2\sigma} \rho^{ab} \pi_{a} \pi_{b},                \label{casimir}
\end{equation}
where the integrand is a scalar density with weight $1$, and $\mathcal{A}$ is the area of  $N_{2}$. The factor  $({\mathcal{A}}/{4\pi} )^{2}$ makes $L^2$ to have the correct dimension of angular momentum squared.

In order to show the equation (\ref{mouse}) is true, 
let us first notice that $L(\xi)$ is the generating function of diffeomorphisms of $N_{2}$
along $\xi^{a}$. This can be seen from the Poisson algebra (\ref{water}), which can be written as 
\begin{eqnarray}
\{ \oint d^{2} y''  \ \xi^{a}(y'') \pi_{a}(y'') , 
\oint d^{2} y' \ \eta^{b}(y') \pi_{b}(y')  \}_{\rm P.B.}
= \oint d^{2} y \ [\xi,\eta]^{a}  \pi_{a} .
\label{red}
\end{eqnarray}
Let us choose the test function $\xi^{a}(y'')$ in (\ref{red}) as
\begin{equation}
\xi^{a}(y'') =  \delta^{(2)} (y''-y) \ \delta^{a}_{j}
\label{yellow}
\end{equation}
for fixed $j$. Then, the equation (\ref{red}) becomes
\begin{eqnarray}
\{ \pi_{j} , \oint d^{2} y' \ \eta^{b}(y') \pi_{b}(y')   \}_{\rm P.B.}
&=& \oint d^{2} y' \ [\xi(y'),\eta(y')]^{a}  \pi_{a}(y')   \nonumber \\
&=& - \oint d^{2} y' \ \pi_{j}(y') \mathcal{L}_{\eta(y')} 
\delta^{(2)} (y'-y)    \nonumber \\
&=& \mathcal{L}_{\eta} \pi_{j} .
\label{green}
\end{eqnarray}
Thus, we find that
\begin{equation}
\{ \pi_{j} ,  L(\eta)  \}_{\rm P.B.}
= {1\over 16\pi} \mathcal{L}_{\eta} \pi_{j} .
\label{cheese}
\end{equation}
Also notice that 
\begin{eqnarray}
\hspace{-0.8in}
\{ \rho^{ij}(y) ,  \oint d^{2} y' \ \eta^{b}(y') \pi_{b}(y')   \}_{\rm P.B.}     
&=& \{ \rho^{ij}(y) , 
\int du' \oint d^{2} y' \pi^{ab}(y') \mathcal{L}_{\eta(y')} \rho_{ab}(y') \}_{\rm P.B.}   \nonumber \\
&=& \int du' \oint d^{2} y' \{ \rho^{ij}(y),  \pi^{ab}(y')  \}_{\rm P.B.}    
\mathcal{L}_{\eta(y')} \rho_{ab}(y') \nonumber \\
&=& \mathcal{L}_{\eta(y)} \rho^{ij}(y),   \label{rain}
\label{blue}
\end{eqnarray}
where we used the equation (\ref{QAM}) in the first line. 
Therefore, it follows that 
\begin{equation}
\{ \rho^{ij} ,    L(\eta) \}_{\rm P.B.}
= {1\over 16\pi}\mathcal{L}_{\eta} \rho^{ij} .
\label{tiguan}
\end{equation}
In a similar way, we find that
\begin{eqnarray}
& & \{ e^{-2\sigma} ,  L(\eta) \}_{\rm P.B.}    
={1\over 16\pi} \mathcal{L}_{\eta} e^{-2\sigma},        \label{choco}\\
& & \{ \mathcal{A} ,  L(\eta)  \}_{\rm P.B.}  
=0.                \label{brown}
\end{eqnarray}
From the equations (\ref{cheese}), (\ref{tiguan}), (\ref{choco}), and (\ref{brown}), we find that $L^{2}$ is invariant under the diffeomorphisms of $N_{2}$,
i.e.,
\be
 \{ L^{2}, L(\eta)  \}_{\rm P.B.}  = 0 .    \label{printer}
\ee 
Thus, $L^{2}$ should be regarded as a quasilocal generalization of the Casimir invariant of ordinary angular momentum in the flat spacetime.

%%%%%%%%%%%%%%%%%%%%%%%%%%%%%%%%%%%%%%%%%%%%%%%%%%%%%%%%%%%%%%%%%%%%%%%%%%%%%%%%%%%%%%%

\section{The asymptotic limit of quasilocal angular momentum} \label{asymptotic}

In this section, the angular momentum of the Kerr spacetime at null infinity will be obtained as the limit of the quasilocal angular momentum discussed in previous sections. Let us assume the integration constant
\be
L_{0}  = 0  \label{backer}
\ee
as before.
At null infinity of asymptotically flat spacetimes, $N_{2}$ becomes $S_2$. 
Let $y^{a} = (\theta, \phi)$, the usual angular coordinates on $S_2$. The generators 
$\bar{\xi}_{(\alpha)} \ (\alpha=1,2,3)$ of the $SO(3)$ isometry of $S_2$ are given by
\bea
\bar{\xi}_{(1)} &=& -\sin \phi \frac{\partial}{\partial \theta} -\cot \theta \cos \phi \frac{\partial}{\partial \phi}, \label{coffee} \\
\bar{\xi}_{(2)} &=& \cos \phi \frac{\partial}{\partial \theta} -\cot \theta \sin \phi \frac{\partial}{\partial \phi}, \label{ice}\\
\bar{\xi}_{(3)} & =&  \frac{\partial}{\partial \phi}, \label{sugar}
\eea
which satisfy the commutation relations 
\bea
[\bar{\xi}_{(\alpha)} , \bar{\xi}_{(\beta)} ]_{\rm L} = -{\epsilon_{\alpha\beta}}^\gamma \bar{\xi}_{(\gamma)}. 
\label{xicom}
\eea
By continuity, there exist tangent vector fields 
$\xi_{(\alpha)} \ (\alpha=1,2,3)$ on $N_2$ in the asymptotic zone of asymptotically flat spacetimes, which approach $\bar{\xi}_{(\alpha)}$ 
as $N_2$ approach $S_2$ in the limit  $v \rightarrow \infty$,
\bea
\lim_{v \to \infty} \xi_{(\alpha)} = \bar{\xi}_{(\alpha)}.
\eea
Let $\bar{L}_\alpha$ be the gravitational angular momentum at null infinity 
defined by the limit
\bea
\bar{L}_\alpha &:=& \lim_{v\to\infty} L(u,v; \xi_{(\alpha)}). \label{gam}
\eea
It is trivial to show that $\bar{\xi}_{(\alpha)}$ is divergence-free,
\bea
\bar{\nabla}_a \bar{\xi}_{(\alpha)}^{a} = 0    \quad (\alpha=1,2,3).       \label{lake}
\eea
Moreover, the following commutation relation is a direct consequence of the theorem proven
in Section \ref{poisson}, 
\bea
\{ \bar{L}_\alpha , \bar{L}_\beta \}_{\rm P.B.} 
= - \frac{1}{16\pi} {\epsilon_{\alpha\beta}}^\gamma \bar{L}_\gamma.  \label{patagonia}
\eea
This shows that the algebra of gravitational angular momentum at null infinity realizes 
the $SO(3)$ Lie algebra under the Poisson bracket.

The angular momentum of the Kerr spacetime can be obtained by taking the limit
(\ref{gam}). Let us notice that the Kerr metric in the asymptotic zone can be written as
\bea
\hspace{-0.5in}
ds^2 = &&-2du dv -(1-\frac{2m}{v} + \cdots ) du^2 + (\frac{4ma \sin^2 \theta}{v}  \nonumber\\
&&-\frac{4ma^3 \sin^2 \theta \cos^2 \theta}{v^3} +\cdots) du d\phi +v^2 (1+\frac{a^2 \cos^2 \theta }{v^2} + \cdots ) d\theta^2        \nonumber\\
&&+v^2 \sin^2 \theta (1+ \frac{a^2}{v^2} + \cdots) d\phi^2 + \sin^2 \theta (\frac{4ma^3}{v^3} + \frac{8m^2 a^3}{v^4} + \cdots ) dv d\phi        \nonumber\\
&& -(\frac{a^2 \sin^2 \theta}{v^2} +\cdots ) dv^2,                    \label{Kerr}
\eea
where $m$ and $a$ are the mass and the specific angular momentum, respectively. 
If we compare the metric (\ref{aborigin}) with the above metric, the leading terms of the metric coefficients are found to be as follows:
\bea
f=1,                                     \label{unitf} \\
e^\sigma = v^2 \sin \theta  + O(1) ,             \label{esigma} \\
\rho_{\theta \theta} = \frac{1}{\sin \theta} + O(1/v),      \label{rhott}                      \\ 
\rho_{\theta \phi} = O(1/v^2),                                        \label{rhotp}                    \\
\rho_{\phi \phi} = \sin \theta + O(1/v),                           \label{rhopp}                   \\
2h = 1 - \frac{2m}{v} + O(1/v^2),                 \\
A_+^\phi = \frac{2ma}{v^3} + O(1/v^4),      \label{aphi+}               \\
A_-^\phi = \frac{2ma^3}{v^5} + O(1/v^6),    \label{aphi-}               \\
A_\pm^\theta = O(1/v^6).                            \label{atheta}
\eea
Inserting (\ref{aphi+}), (\ref{aphi-}), and (\ref{atheta}) into the equation (\ref{F+-}), we find
that
\bea
F_{+-}^\theta = \partial_+ A_-^\theta - \partial_- A_+^\theta  - A_+^a \partial_a A_-^\theta + A_-^a \partial_a A_+^\theta = O(1/v^6),              \label{ftheta} \\
F_{+-}^\phi = \partial_+ A_-^\phi - \partial_- A_+^\phi  - A_+^a \partial_a A_-^\phi + A_-^a \partial_a A_+^\phi = \frac{6ma}{v^4} + O(1/v^5).             \label{fphi} 
\eea
Let us choose the arbitrary constant $k=1$ so that the normalization condition
becomes
\bea
<n,l> \ = -1.
\eea
Then, by the equation (\ref{transnl}), 
the tangential components $[n,l]_{\rm L}^{a}$ are
given by
\bea
[n,l]_{\rm L}^{a} = - F_{+-}^{a},  \label{tan}
\eea
from which it follows that
\bea
&&[n,l]_{\rm L}^\theta  \longrightarrow  O(1/v^6),                    \label{nltheta}      \\
&&[n,l]_{\rm L}^\phi \longrightarrow  - \frac{6ma}{v^4} + O(1/v^5).         \label{nlphi}
\eea
After straightforward calculations, we find that the asymptotic 
limits of the integrands of the equation (\ref{ensenble}) for $\xi_{(\alpha)} (\alpha = 1,2,3) $ are given by
\bea
e^\sigma <\xi_{(1)}, [n,l]_{\rm L}> \longrightarrow 6ma\sin^2 \theta \cos \theta \cos \phi + O(1/v),            \label{xi1} \\
e^\sigma <\xi_{(2)}, [n,l]_{\rm L}> \longrightarrow 6ma\sin^2 \theta \cos \theta \sin \phi + O(1/v),             \label{xi2} \\
e^\sigma <\xi_{(3)}, [n,l]_{\rm L}> \longrightarrow -6ma\sin^3 \theta + O(1/v).                                        \label{xi3} 
\eea
If we plug (\ref{xi1}), (\ref{xi2}), and (\ref{xi3}) into (\ref{ensenble}), we find that
\bea
L(u,v;\xi_{(1)})
&=&-\frac{1}{16\pi} \int_0^\pi d\theta  \int_0^{2\pi} d\phi~\big( 6ma\sin^2 \theta \cos \theta \cos \phi \big)+ O(1/v)       \nonumber \\
&=& O(1/v),                                                                                                                                                                                                   \label{ql1}                 \\
L(u,v;\xi_{(2)}) 
&=&-\frac{1}{16\pi} \int_0^\pi d\theta  \int_0^{2\pi} d\phi~\big( 6ma\sin^2 \theta \cos \theta \sin \phi \big)+ O(1/v)       \nonumber \\
&=& O(1/v),                                                                                                                                                                                                    \label{ql2}                 \\
L(u,v;\xi_{(3)}) 
&=&\frac{1}{16\pi} \int_0^\pi d\theta  \int_0^{2\pi} d\phi~\big( 6ma\sin^3 \theta \big)+ O(1/v)       \nonumber \\
&=& 
ma + O(1/v),  \label{q13}                                                                                                                                  
\eea
respectively. In the limit  $v \rightarrow \infty$, they become
\bea
\bar{L}_1 &=& \lim_{v\to\infty} L(u,v; \xi_{(1)} ) =0 ,                                                                                                              \label{l1}\\
\bar{L}_2 &=& \lim_{v\to\infty} L(u,v; \xi_{(2)} ) =0 ,                                                                                                              \label{l2}\\
\bar{L}_3 &=& \lim_{v\to\infty} L(u,v; \xi_{(3)} ) =ma .                                                                                                            \label{l3}
\eea
Thus, the only non-vanishing component of the angular momentum is along the
vector field
\begin{equation} 
\lim_{v\to\infty} \xi_{(3)} = \bar{\xi}_{(3)} =  \frac{\partial}{\partial \phi},
\end{equation}
and reproduces the angular momentum $ma$ of the Kerr spacetime at null infinity.

Finally, let us show that the invariant quasilocal angular momentum 
$L^{2}$ defined in (\ref{casimir}) becomes $(ma)^{2}$ in the same limit.
If we use the equation (\ref{pia}) and the asymptotic expansions for the fields
with $f=1$, then we find that
\begin{equation}
 e^{-2 \sigma} \rho^{ab} \pi_{a} \pi_{b} 
 = \frac{36 (ma)^{2}}{v^{4}} \sin^{3} \theta 
 + O(\frac{1}{v^{5}}).
\label{expansion}
\end{equation}
Thus, we have 
\begin{equation}
\lim_{v \to \infty} L^2 = (ma)^{2}  . 
\label{square}
\end{equation}

%%%%%%%%%%%%%%%%%%%%%%%%%%%%%%%%%%%%%%%%%%%%%%%%%%%%%%%%%%%%%%%%%%%%%%%%%%%%%%%%%%%%%%%%

\section{Rizzi's geometric angular momentum as the limit of quasilocal angular momentum}\label{rizzi}

In this section, we will argue that the quasilocal angular momentum (\ref{ensenble}) is 
a natural generalization of the angular momentum proposed by A. Rizzi to a finite distance.
Rizzi's approach is based on an affine foliation of the asymptotic zone of a asymptotically flat spacetime with families of $S_2$ \cite{rizzi98}. 
He introduced the in- and out-going null vector fields  
$n$ and $l$ with the normalization
\bea
<n , l > =-2.            \label{norm}
\eea
Together with an orthonormal frame $e_a$ ($a=1,2$) on $S_2$, $n$ and $l$ form a tetrad system 
\bea
\{~ e_a, e_3, e_4 : a=1,2,~e_3=n, e_4= l ~\}.
\eea
The twist $\zeta_a$ of $n$ and $l$ is defined as
\bea
\zeta_a=\frac{1}{2} <\nabla_3 e_4 , e_a>.                \label{twist}
\eea
It can be written as
\bea
\zeta_a &=& \frac{1}{4} (<\nabla_3 e_4 , e_a>-<\nabla_4 e_3 , e_a>) \nonumber\\
&=&\frac{1}{4}<[e_3, e_4]_{\rm L}, e_a> \nonumber\\
&=&  \frac{1}{4} [n,~l]_{{\rm L}a} .           \label{zeta}
\eea
Rizzi's angular momentum is defined as the integral
\bea
L(\Omega_{(\alpha)}) = -\frac{1}{8\pi} \lim_{s\to\infty} \oint_{S_2} \Omega_{(\alpha)}^a \zeta_a  dS_\gamma ,                            \label{def}
\eea
where $\Omega_{(\alpha)}$ are the SO(3) generators that satisfy the commutation relations
\bea
[\Omega_{(\alpha)},\Omega_{(\beta)}]_{\rm L} 
= -\epsilon_{\alpha\beta}^{\ \ \gamma}\Omega_{(\gamma)},              \label{sign}
\eea
and $s$ is an affine parameter of out-going null vector field $l$
(Notice the change of sign from the Rizzi's original definition).
The right hand side of (\ref{def}) involves the limit that pulls the variables back to $S_2$, 
and $dS_\gamma$ is the infinitesimal area element of $S_2$ with the standard metric $\gamma$.  
If we use $\zeta_a$ given by (\ref{zeta}), then Rizzi's angular momentum becomes
\bea
L(\Omega_{(\alpha)})
& =& -\frac{1}{32\pi} \lim_{s\to\infty} \oint_{S_2} \Omega_{(\alpha)}^a [n,~l]_{{\rm L}a} 
 dS_\gamma \nonumber\\
&=&\frac{1}{16\pi <n,l>} \lim_{s\to\infty} \oint_{S_2} 
\Omega_{(\alpha)}^a [n,~l]_{{\rm L}a}
dS_\gamma,                  \label{Rizzi}
\eea
which is precisely the limit (\ref{gam}) of our quasilocal angular momentum
(\ref{ensenble})
with the following identifications
\begin{equation}
L_{0}=0, \quad dS_\gamma = d \mu,   \quad \Omega_{(\alpha)}^{a} =\bar{\xi}_{(\alpha)}^{a}.
\end{equation}

%%%%%%%%%%%%%%%%%%%%%%%%%%%%%%%%%%%%%%%%%%%%%%%%%%%%%%%%%%%%%%%%%%%%%%%%%%%%%%%%%%%%%%%

\section{Discussion}\label{discussion}

It is remarkable that two-surface integrals of the Einstein's constraint equations in the (2+2) Hamiltonian formalism are split into terms that can be geometrically interpreted as quasilocal energy, linear momentum, angular momentum on the two-surface and 
the corresponding fluxes crossing that surface. In particular, we have shown that the Poisson algebra of the quasilocal angular momentum associated with an arbitrary tangent vector field $\xi$ on $N_{2}$ is closed,
\bea
\{ L(\xi), L(\eta) \}_{\rm P.B.} = \frac{1}{16\pi} L([\xi, \eta]_{\rm L}).  \label{alcohol}
\eea
Furthermore, we showed that the necessary and sufficient condition that ensures the gauge-independence of 
the quasilocal angular momentum is 
\begin{equation}
\tilde{\nabla}_{a} \xi^{a} = 0,
\end{equation}
where $\tilde{\nabla}_{a}$ is the covariant derivative on $N_{2}$. The divergence-free vector field  $\xi$ 
need not be a Killing vector field on $N_{2}$, and therefore the quasilocal angular momentum  $L(\xi)$ is well-defined even in the absence of any isometries. 
However, as was discussed in Section  \ref{invariant}, any continuous tangent vector field has at least one zero for any 2-dimensional compact manifold with non-zero Euler characteristic, so that $L(\xi )$ fails to measure the rotation of gravitational fields at the singular points of $\xi^{a}$ on such $N_{2}$. 
This observation led us to introduce
the invariant quasilocal angular momentum $L^{2}$ of gravitational fields on a given 
$N_{2}$, which is free from the singularity of the divergence-free tangent vector field $\xi^{a}$
and commutes with  $L(\xi)$,
\be
 \{ L^{2}, L(\xi)  \}_{\rm P.B.}  = 0,  \label{rat}
\ee 
and becomes
\begin{equation}
\lim_{v \to \infty} L^2 = (ma)^{2} \          \label{triangle}
\end{equation}
at the null infinity of the Kerr spacetime. Therefore, we conjectured the invariant quasilocal angular momentum $L^{2}$ given by (\ref{casimir}) as a gravitational analog of the Casimir invariant of ordinary angular momentum.

We also showed that the asymptotic limit of our quasilocal angular momentum is identical to the geometric angular momentum of A. Rizzi at null infinity, and that it satisfies the $SO(3)$ commutation relations
\bea
\{ \bar{L}_\alpha , \bar{L}_\beta \}_{\rm P.B.} = -\frac{1}{16\pi} {\epsilon_{\alpha\beta}}^\gamma \bar{L}_\gamma.
\eea
In spite of the agreement of the two definitions in the asymptotic limit, there are differences that must be mentioned. 
Firstly, Rizzi's angular momentum requires the affine foliation 
of spacetimes at null infinity, whereas our angular momentum does not even at a finite distance. Our quasilocal angular momentum is based on the (2+2) fibre bundle decomposition
of spacetimes, where in- and out-going null vector fields need not be affinely parametrized
at all. 

Secondly, our quasilocal angular momentum, though obtained in a canonical formalism, 
admits a natural geometrical interpretation, whereas Rizzi's definition is purely geometrical that lacks a canonical interpretation. Moreover, by the Einstein's constraint equation, our quasilocal angular momentum can be also represented as a 3-dimensional integral of flux, as is shown in (\ref{QAM}). Using this
3-dimensional integral representation of quasilocal angular momentum, 
we were able to show that the Poisson algebra of our quasilocal angular momentum 
reduces to the $SO(3)$ algebra at the null infinity.
It would be hard to obtain this algebra from Rizzi's definition even at null infinity,
due to the lack of the canonical representation.

Thirdly, unlike Rizzi's definition, our quasilocal angular momentum and its flux integrals follow directly from the  Einstein's constraints. However, 
it must be also said that 
the splitting of the Einstein's constraint into angular momentum and its flux part is not unique;
one can always add an arbitrary quantity to each part without changing the constraint equation as a whole. By the criteria of a satisfactory definition of quasilocal angular momentum stated in Section \ref{intro}, however, one can reasonably eliminate such ambiguities. Thus, our expressions of quasilocal angular momentum and its flux are fairly unique.

Finally, one might try to find the Poisson algebra and its limit of quasilocal gravitational energy, linear and angular momentum, and see whether they form the Poincar{\`e} 
algebra or its BMS generalization in the asymptotic limit. This problem is interesting in its own right, and will be reported in a forthcoming paper. 

%%%%%%%%%%%%%%%%%%%%%%%%%%%%%%%%%%%%%%%%%%%%%%%%%%%%%%%%%%%%%%%%%%%%%%%%%%%%%%%%%%%%%%%

\ack{The authors would like to express deep thanks to Prof. L. Szabados who suggested to check the gauge-independence of the quasilocal angular momentum by the method in Section 3 of this paper, and Prof. D. R. Brill who suggested the authors to look for the invariant quasilocal angular momentum at the 13th International Conference on Gravitation, Astrophysics, and Cosmology (Seoul, Korea 2017). This work was supported by National Research Foundation of Korea (Grant 2015-R1D1A1A01-059407).}

%%%%%%%%%%%%%%%%%%%%%%%%%%%%%%%%%%%%%%%%%%%%%%%%%%%%%%%%%%%%%%%%%%%%%%%%%%%%%%%%%%%%%%%%

\end{document}